\documentclass[a4paper]{article}
\usepackage[margin=25mm]{geometry}
\usepackage{amsmath}
\usepackage{amsfonts}
\usepackage{amssymb}
\usepackage{graphicx}
\usepackage[colorlinks=true, allcolors=blue]{hyperref}
\usepackage{verbatim}
\immediate\write18{texcount -tex -sum  \jobname.tex > \jobname.wordcount.tex}

\usepackage[font=tiny,labelfont=bf, figurename=Fig.]{caption}
\providecommand{\keywords}[1]
{
  \small	
  \textbf{\textit{Keywords---}} #1
}

\title{The cost of artificial latency in the PBS context}

\author{Umberto Natale\footnote{Contact Author}\\Chorus One\\umberto@chorus.one \and Michael Moser\\Chorus One\\michael.moser@chorus.one}

\begin{document}
\maketitle

\begin{abstract}
\noindent
We present a comprehensive analysis of the implications of artificial latency in the Proposer-Builder Separation framework on the Ethereum network. Focusing on the MEV-Boost auction system, we analyze how strategic latency manipulation affects Maximum Extractable Value yields and network integrity. Our findings reveal both increased profitability for node operators and significant systemic challenges, including heightened network inefficiencies and centralization risks. We empirically validates these insights with a pilot that Chorus One has been operating on Ethereum mainnet. We demonstrate the nuanced effects of latency on bid selection and validator dynamics. Ultimately, this research underscores the need for balanced strategies that optimize Maximum Extractable Value capture while preserving the Ethereum network's decentralization ethos.

\end{abstract}


\section{Introduction}

In music, a piece played ``adagio'' is performed slowly, and with great expression. This concept extends to an MEV-Boost \cite{mevboost_specs} pilot that Chorus One \cite{c1web} has been operating on Ethereum mainnet. Our setup strategically delays getHeader requests to maximize Maximum Extractable Value (MEV) capture; correspondingly, we've dubbed it \textit{Adagio}. We are committed to operational honesty and rational competition. This study will disclose our results and initial parameters, alongside an extensive discussion of node operator incentives and potential adverse knock-on effects on the Ethereum network.

We argue that strategic latency optimization is a zero sum game which may increasingly compromise node operator independence in favor of statistical arbitrageurs. The practical consequence is centralization pressure over the entire Proposer-Builder Separation (PBS) value chain, leading to ongoing and escalating costs for Ethereum users, and ultimately node operators themselves. This effect is likely to perpetuate as a self-reinforcing cycle, and favors large node operators systematically through reduced payoff variance and a data edge that scales with voting power, leading to centralization. 

The goal of the article is to address and mitigate these competitive dynamics by providing an extensive analysis informed by proprietary data from our Adagio study. Our primary objective is to describe the auction dynamics that give rise to latency strategies, and the associated externalities imposed on the Ethereum network. Our secondary objective is to provide practical context through a discussion of the Adagio setup, and node operator incentives.

\textbf{PBS inefficiencies and MEV returns} is discussed in Sec. \ref{sec:pbs_dyn}. The concept of MEV is intrinsically linked to time via the inherent volatility of asset prices. In the current PBS landscape, there exists an inefficiency with regards to auction timing, which can be strategically exploited to consistently generate excess MEV returns \cite{time_is_money}. Our study highlights this inefficiency by examining the MEV-Boost auction’s bid dynamics in the context of timing games. 

\textbf{The cost of artificial latency} is discussed in Sec. \ref{sec:cost_of_latency}. We examine how latency optimization generates excess MEV rewards at the expense of subsequent proposers, while increasing the Ethereum burn rate, contributing to gas cost. Further, we discuss how a delayed auction imposes additional Loss-Versus-Rebalancing (LVR) \cite{LVR} burden on liquidity providers (LPs) by exacerbating the disconnect between on-chain asset prices and other venues, while decreasing inventory risk for statistical arbitrageurs. We argue that all of the above contributes to node operator centralization, and that not engaging in timing games imposes an excess opportunity cost on node operators of any voting power. 

\textbf{Empirical results from the Adagio pilot} are discussed in Sec. \ref{sec:adagio_results}. We present our experimental setup and benchmark its performance against the network. Further, the chapter discusses the idiosyncrasies of optimistic- and non-optimistic relays in the context of timing games. We close with a statistical analysis of the additional MEV revenue captured by Adagio versus a non-optimized setup. 

Through this structured approach, we aim to provide a comprehensive overview of the effects of latency optimization within the PBS framework, highlighting the opportunities and challenges it presents to the Ethereum ecosystem.

\section{PBS dynamics, and the MEV-Boost auction}\label{sec:pbs_dyn}

The Ethereum network seeks to maximize decentralization by allowing hobbyists to run validators at a competitive cost / benefit ratio. Proposer-Builder Separation (PBS) \cite{pbs_specs} enables this by separating the block proposal and construction roles, with builders competing to provide maximally valuable blocks to proposers. Complexity is pushed downstream to sophisticated builders, whereas validators participating in consensus can access optimized blocks featuring Maximum Extractable Value (MEV)\cite{MEV} transactions, irrespective of their degree of sophistication.

MEV-Boost works through a commit and reveal scheme where proposers commit to blocks created by builders without seeing their contents, by signing block headers. Only after a block header is signed, the block body and corresponding transactions are revealed. A trusted third party called a relay facilitates this process. This mechanism ensures that validators cannot steal
the content of a builder’s block, but can still commit to proposing said block by signing the corresponding header. 

The MEV-Boost block auction initiates one slot ahead of the block's intended slot. Builders submit block proposals and bids to relays to determine who gets to construct the next block and capture MEV profits. Relays log when each bid is received (\textbf{receiving time}) and, after checks, make them available to the proposer (\textbf{eligibility time}). The proposer then requests the best bid's block header (\textbf{getHeader request}) from the relays, checks for the best bid, signs it to commit to the block, and asks the relay for the full block content (\textbf{getPayload} and \textbf{submitBlindedBlock}). Once the relay receives the signed header, it broadcasts the full block to the network -- cfr. Fig. 1 of \cite{time_is_money}. After a delay \cite{lowcarbcruc}, the relay sends the payload to the proposer.

The subsequent section will examine the auction dynamics inherent to the MEV-Boost architecture, expanding on builder competition and the resultant network behaviors. It is important to note that in our analysis, we often refer to the \textbf{receiving time} of a bid \cite{bidSubRelay} as its \textbf{eligibility time}. This conflation is intentional and based on the close correlation between the two, with the eligibility time typically lagging the receiving time by an estimated 100 to 200 milliseconds \cite{optRel}. While relays do not explicitly provide eligibility timestamps, our references to eligibility time are grounded on the logical inference of this correlation.

The insights presented here are not static; they contribute to the ongoing discourse surrounding Ethereum's path towards a more equitable and robust, incentive compatible setup.

\subsection{PBS auction dynamics}\label{subsec:pbs_auct}

\begin{figure}[h!]
  \centering
  \includegraphics[width=0.8\textwidth]{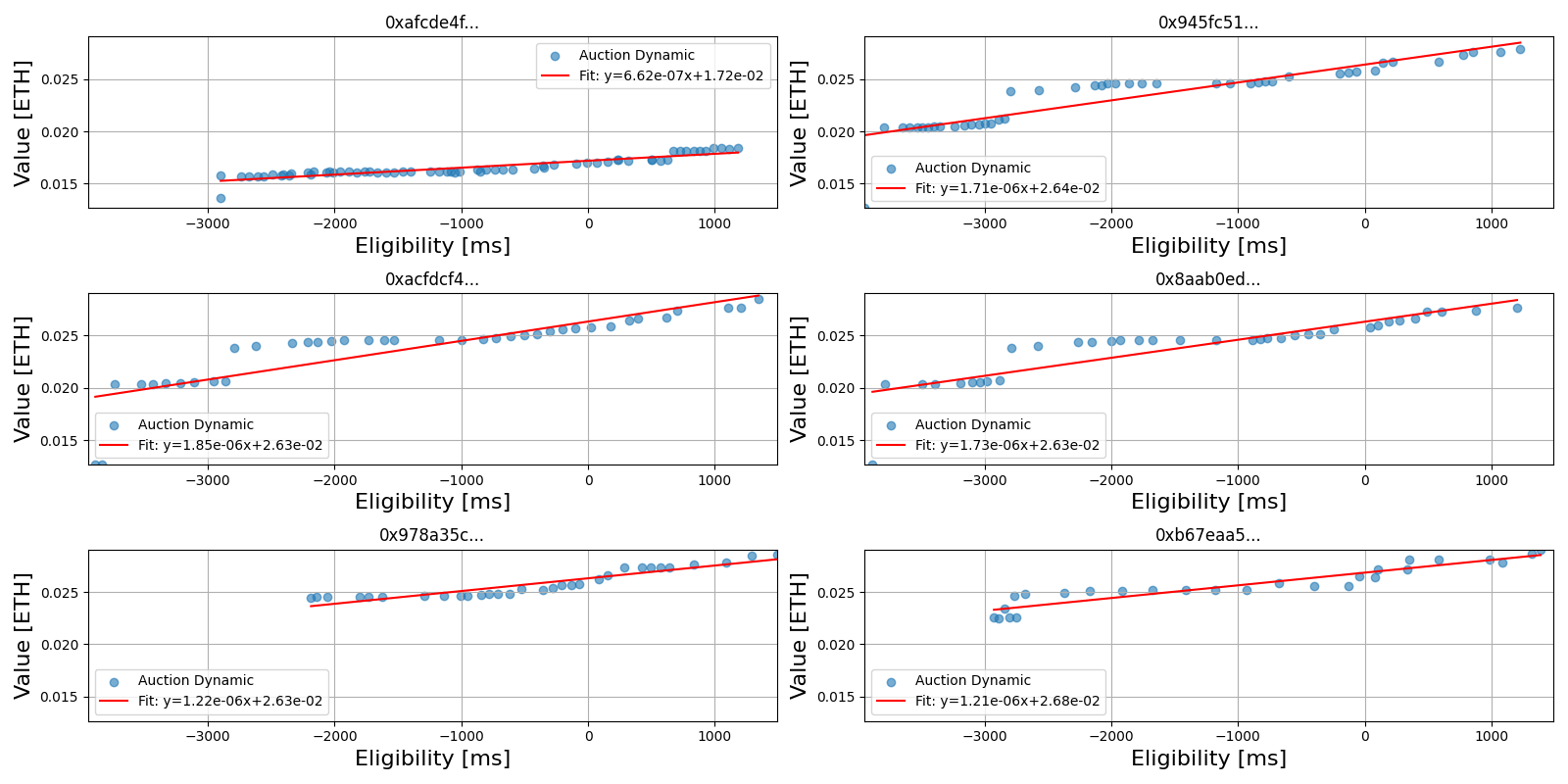}
  \caption{Representation of bid value evolution as a function of time into the slot for different builders account. Bid values are represented with dots, where the red line stands for the linear regression fit.}
  \label{fig:builders_competition}
\end{figure}

\noindent
The competitive dynamics between builders give rise to the MEV-Boost auction. Notably, a given builder's bid performance is not solely a function of its capital and computational efficiency, but of informational reach. Figure \ref{fig:builders_competition} illustrates this effect for a single slot and relay; the pattern extends to the general case (i.e. all builders, relays, and slots). Here, builder \texttt{0xafcde4f...} (pictured top left) undercompetes due to lack of information vs. builders \texttt{0x945fc51...}, \texttt{0xacfdcf4...}, \texttt{0x8aab0ed...}, and \texttt{0x978a35c...}. All of the latter are associated with the same entity: rsync (cfr. \cite{MEVBoostDash}). The last builder featured in the graph, \texttt{0xb67eaa5...}, is Titan, a highly competitive neutral builder. While the prototypical example of an information edge is private order flow, the definition extends to capture other systematic edges that can increase bids, e.g. a proprietary statistical arbitrage model for an integrated builder-searcher entity. Ultimately, information manifests as bargaining power.

The structure we see on a granular scale -- i.e. per slot and per builder -- manifests also on the aggregate level -- i.e. considering all slots and all builders. To allow for a standardized comparison of bids irrespective of their absolute size, we define the following normalized estimator as the ratio between a given bid and the maximum bid in the auction. That is
\begin{equation}\label{eq:Restimator}
R = \frac{b(t_E)}{\max_{t_E \le 2\,s} b}\,,
\end{equation}
\noindent
where $b$ is the bid value, and $t_E$ is the corresponding time at which the bid was made eligible. Here we use the 2 s mark since there is no evidence for the network going above this limit, cfr. \cite{data_always_bid_time, MEVBoostDash}.

Consider a scenario where the winning bid is 700 ETH. If, after 100 milliseconds, the relay makes a higher bid of 770 ETH eligible, this would conventionally represent a 10\% increase. In our normalized system, the $R$ value for the initial winning bid when compared to the maximum of 770 ETH would be approximately $R=\frac{700}{770}\sim 0.9$. Conversely, if the winning bid is only 0.1 ETH and a subsequent bid of 0.11 ETH is made eligible after the same time interval, the increase is proportionally identical, and the $R$ value remains the same at $R = \frac{0.1}{0.11}\sim0.9$.

\begin{figure}[h!]
  \centering
  \includegraphics[width=0.8\textwidth]{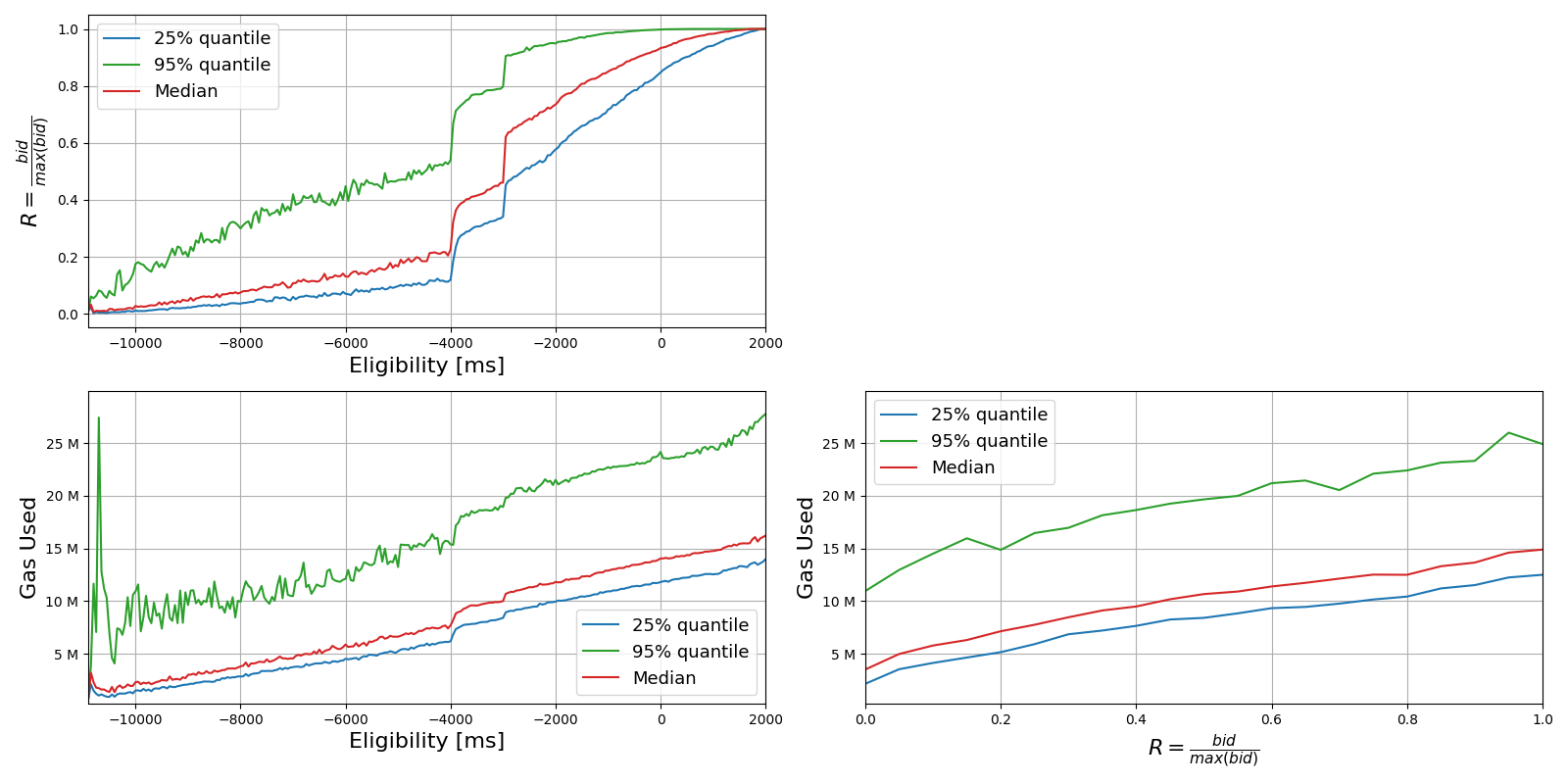}
  \caption{The top panel shows how the bid value compared to the maximum bid in the auction evolves with time into the slot (removing bids made eligible after 2 s). The lower left panel shows how gas used to build the block evolves as a function of eligibility. The lower right panel shows how gas used relates to bid value appreciation. For all panels, the red line represents the median of the distribution, the blue line represents the 25\%-quantile, and the green line represents the 95\%-quantile.}
  \label{fig:bid_value_vs_gas_vs_eligibility}
\end{figure}

Figure \ref{fig:bid_value_vs_gas_vs_eligibility} illustrates the relationship between bid values, gas usage, and bid eligibility over time within the MEV-Boost auction environment. The visualization is divided into three panels, each providing a distinct perspective on the auction's dynamics. In the first panel, the $R$ value, defined as the ratio of a bid to the maximum bid within a slot, is plotted against the eligibility time in milliseconds. This panel highlights the trend that as the eligibility time approaches the 2-second mark, the $R$ value increases, indicating that bids made closer to the 2-second cutoff tend to be larger relative to the maximum bid of the slot. 
Notably, we discern two distinct phase transitions around -4 seconds and -3 seconds into the slot, where the bid value sharply increases -- changing the slope of the bid value increase -- suggesting the flow of new information into the strategies implemented by builders. The second panel shows the gas used in the Ethereum network as a function of eligibility time. This graph shows an upward trend in gas usage as the eligibility time increases, suggesting that bids made later in the slot consume more gas. The third panel correlates the $R$ value directly with the gas used, illustrating how the proportionate size of a bid compared to the maximum bid impacts the amount of gas consumed. We see that $R$ values are generally associated with higher gas usage, reinforcing the idea that blocks which consume more gas are more valuable. Regardless of timing games, we emphasize how the resource use of Ethereum always be on average 15M gas used per slot, due to the EIP-1559 controller dynamics, see e.g. \cite{optFeeMarket}. However, the impact on for the next proposer might have meaningful consequences.

\begin{figure}[h!]
  \centering
  \includegraphics[width=0.8\textwidth]{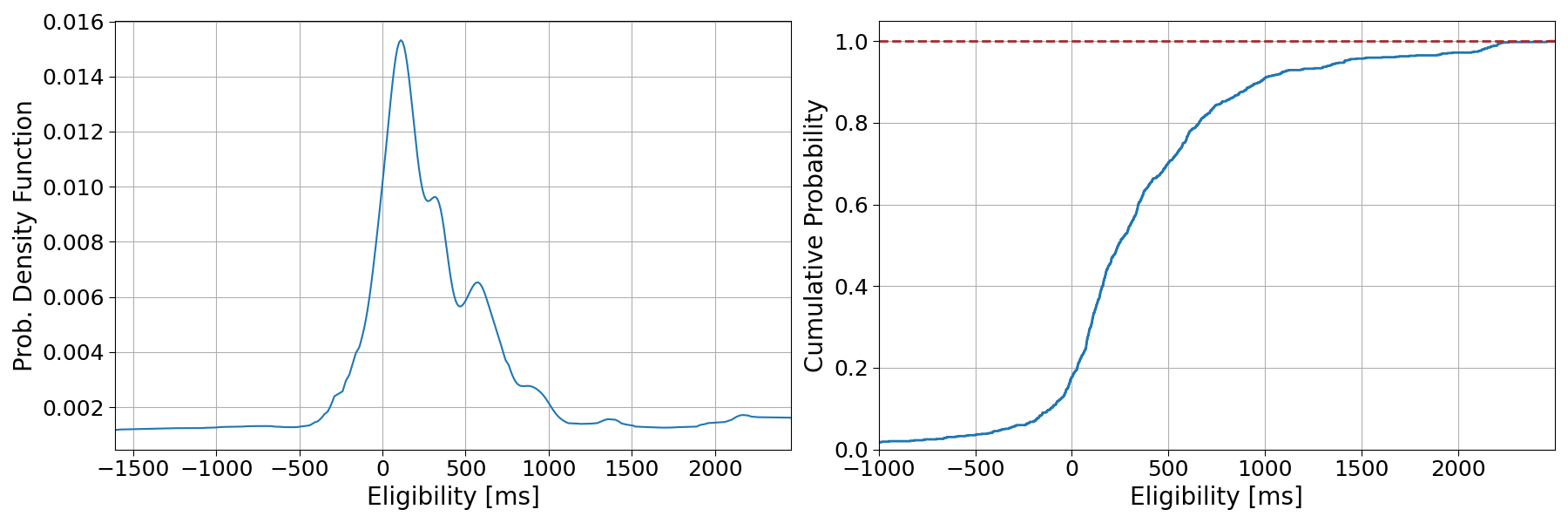}
  \caption{(Left panel) Probability density function of the eligibility time of the winning bids for the sample used in this analysis. (Right panel) Cumulative probability of eligibility times for winning bids.}
  \label{fig:eligibility_pdf_and_cumprob}
\end{figure}

Figure \ref{fig:eligibility_pdf_and_cumprob} depicts the probability density function (PDF) and the cumulative probability of eligibility times for winning bids within our analyzed sample. The 25\%-quantile, median, and 95\%-quantile eligibility times are respectively 74 ms, 240.5 ms, and 1,410m s. Despite the smaller sample size compared to \cite{time_is_money}, our findings regarding the median eligibility time are mutually consistent, corroborating the 260 ms eligibility figure reported in \cite{time_is_money}.

\begin{figure}[h!]
  \centering
  \includegraphics[width=0.8\textwidth]{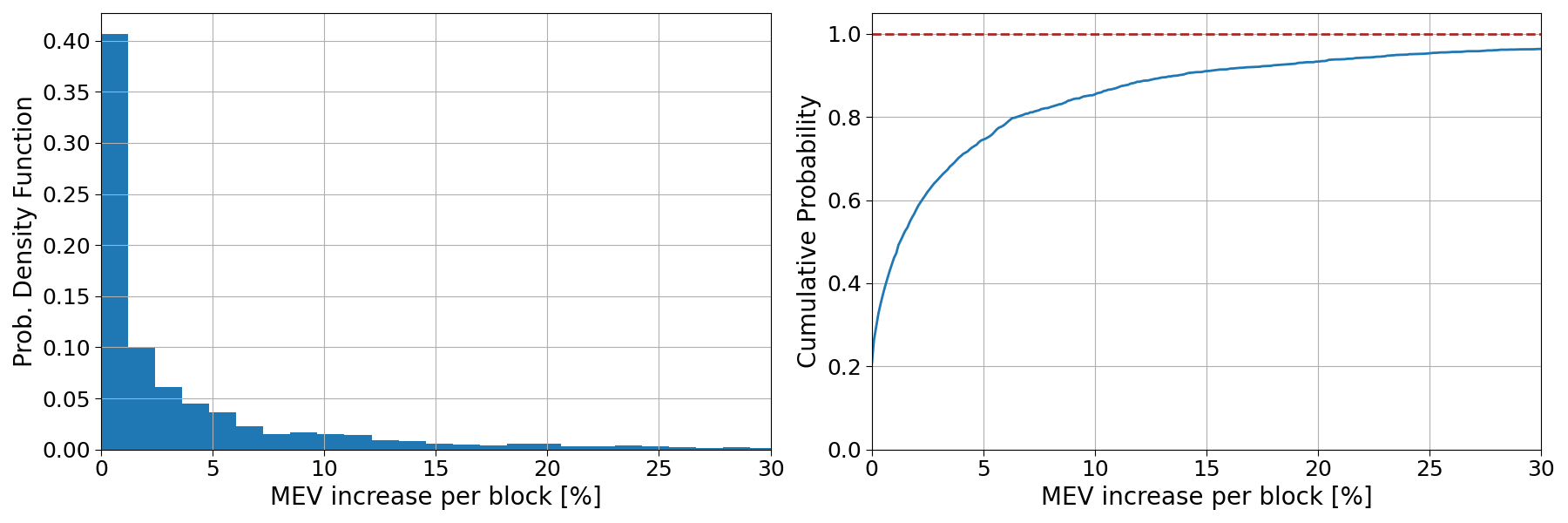}
  \caption{(Left panel) Probability density function of MEV increases per block. (Right panel) Cumulative probability of MEV increases per block.}
  \label{fig:increase_pdf_and_cumprob_real}
\end{figure}

Figure \ref{fig:bid_value_vs_gas_vs_eligibility} and \ref{fig:eligibility_pdf_and_cumprob} indicate that rational validators acting within the bounds of protocol honesty may capture additional MEV by deliberately delaying the auction. Figure \ref{fig:increase_pdf_and_cumprob_real} extends our analysis to the payoff validators may realize by introducing artificial latency.
On the left, the probability density function skews towards the lower end of the MEV increase spectrum, suggesting moderate increases are more common than relatively larger one.
The 25\%-quantile, median, and 95\%-quantile are 0.07\%, 1.28\%, and 23.99\%, respectively. On the right, the cumulative probability function reflects this by showing a rapid initial rise that tapers off. The curve demonstrates that substantial MEV increases are possible, but statistically less likely.

The data depicted in Fig. \ref{fig:increase_pdf_and_cumprob_real} has been synthesized via 1,000 simulations. Each run extracts an eligiblity time from the cumulative distribution in Fig. \ref{fig:eligibility_pdf_and_cumprob}, and computes how much additional MEV could have been extracted if eligibility had been delayed up to 950 ms into the slot. This methodology ensures that our analysis reflects the empirical dynamics of bid eligibility times. The threshold of 950 ms is used because we believe it is a safe region for sampling, since we believe it reduces both the chance of missed slots and the number of forks. It is worth mentioning that this parameter constitutes a heuristic, and further research is needed.

Going forward, we use a standard delay of 950 ms unless otherwise indicated. Exceeding this limit would mean slowing down the auction beyond one second, which could result in an increase in forks, decreasing network efficiency. For operators, the relative decrease in the slope of the $R$ value in the final phase of the auction alongside a relative increase in the risk of missing the slot shifts the risk / reward ratio of artificial latency parameters outside the sampling region unfavorably.

\subsection{Expected node operator return through latency optimization}

We proceed by establishing an empirical framework for potential annual excess yields that can accrue to validators optimizing for latency. This will extend our examination of auction dynamics, and take into account factors including the frequency and scale of MEV opportunities, network conditions, and different latency strategies. To do so, this section will step through three independent variables: MEV rewards per block, the number of blocks a validator is expected to propose, and the MEV increase per block due to latency optimization. In this way, we'll arrive at a robust view of the additional APR can that be realized by small- and large- latency-aware node operators, i.e. their magnitude of incentives.

\begin{figure}[h!]
  \centering
  \includegraphics[width=0.8\textwidth]{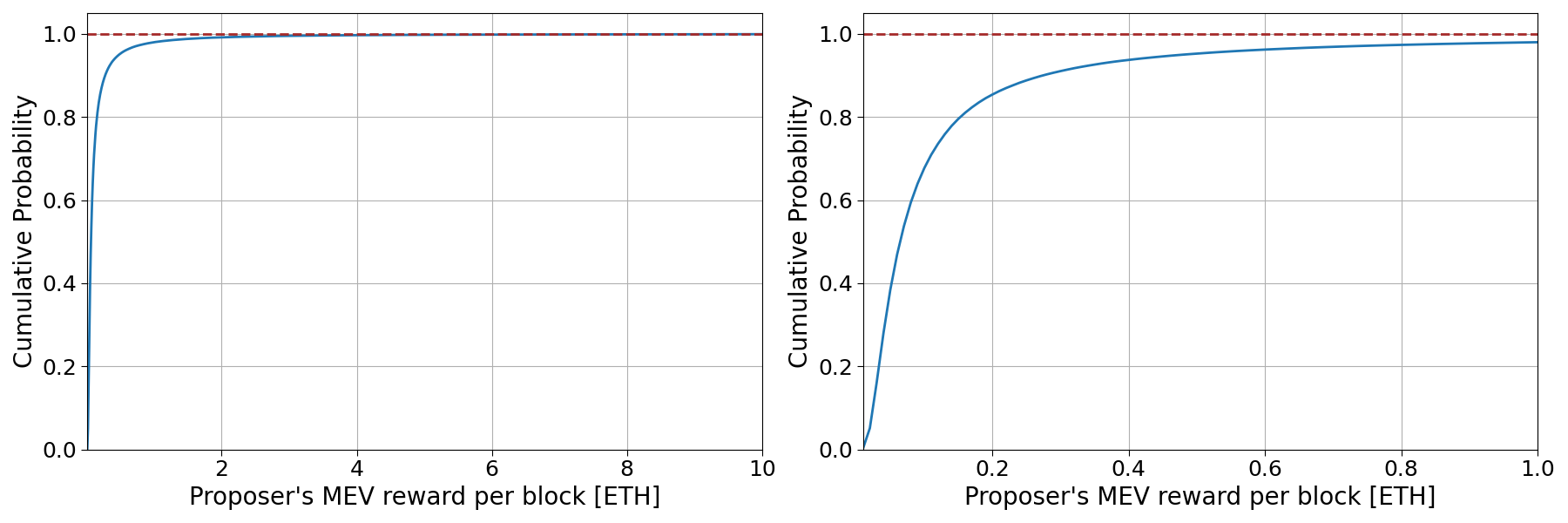}
  \caption{(Left panel) Cumulative probability of proposer’s MEV rewards per block from 2023.01.01 to 2023.11.27. (Right panel) zoom of the left panel in the range 0-1 ETH.}
  \label{fig:MEV_cumprob_block}
\end{figure}

\begin{figure}[h!]
  \centering
  \includegraphics[width=0.8\textwidth]{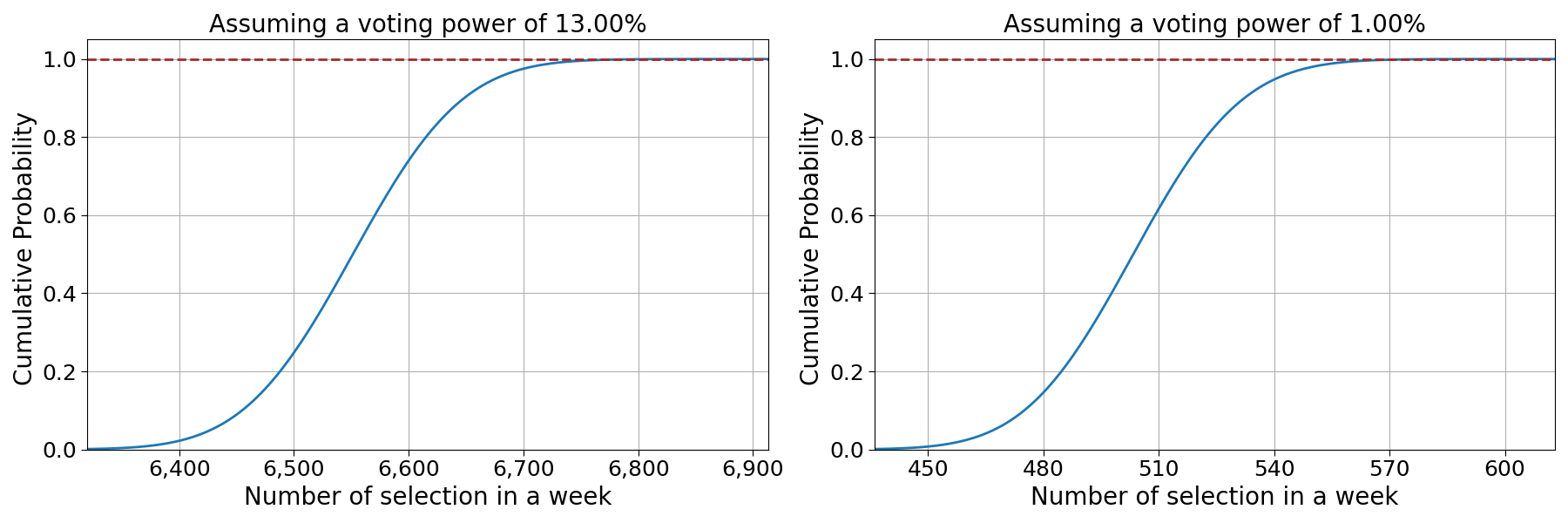}
  \caption{Cumulative probability for number of selection as block proposer in a week for a node operator with 13\% of voting power (left panel) and 1\% of voting power (right panel).}
  \label{fig:prop_selection_cumprob_week}
\end{figure}

First, we will assess the likely absolute per-block MEV payoff. Fig. \ref{fig:MEV_cumprob_block} shows the cumulative probability of MEV rewards per block, based on actual payload data provided to proposers from January 1, 2023, to November 27, 2023 \cite{MEVhealtDash}. The steep ascent of the curve at the lower end of the MEV scale demonstrates that the substantial majority of blocks yields a comparatively small reward. In practical terms, rare, large opportunities make up a disproportionate share of aggregate income. Next, to profit from MEV, a node operator needs to be selected as block proposer. Fig. \ref{fig:prop_selection_cumprob_week} illustrates the cumulative probability distribution for block proposer selection in a week, for node operators with two distinct levels of voting power. For a node operator with a total 13\% voting power, the expected number of block proposals in a week is 6,552. This decreases to 504 weekly opportunities for a node operator with a total 1\% voting power.

\begin{figure}[h!]
  \centering
  \includegraphics[width=0.8\textwidth]{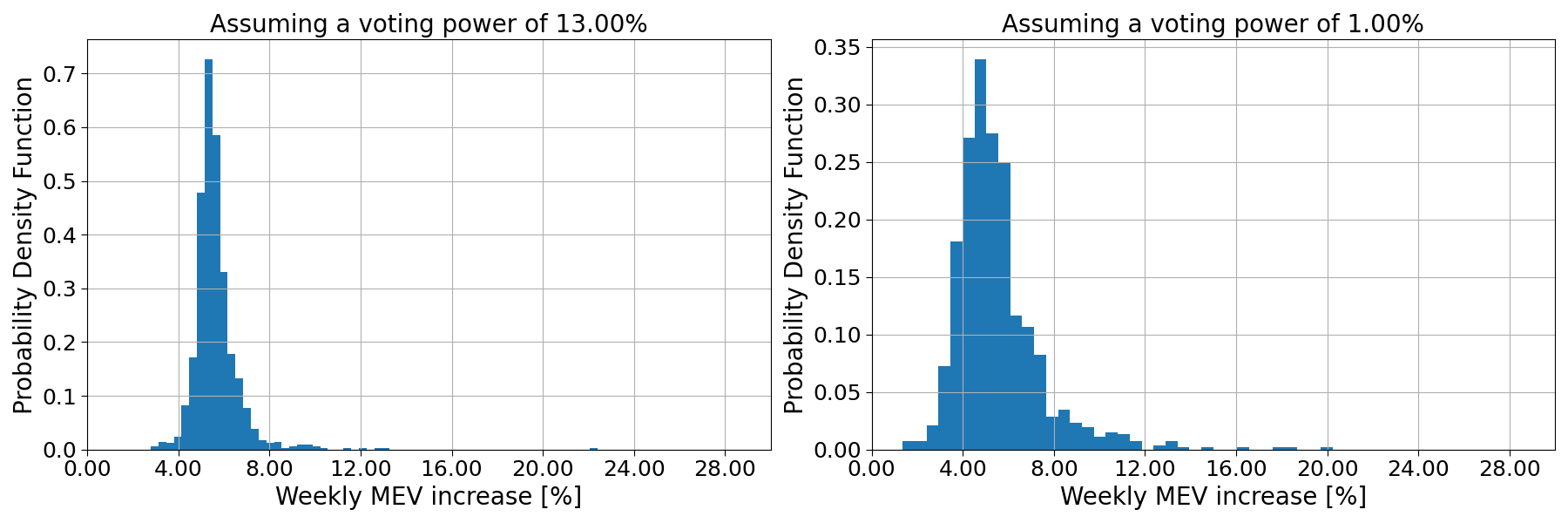}
  \caption{PDF of weekly MEV increase expected by a node operator with 13\% voting power (left panel) and 1\% voting power (right panel).}
  \label{fig:mev_inc_week_vp}
\end{figure}

Combining the latency optimization payoff (Fig. \ref{fig:increase_pdf_and_cumprob_real}), per-block MEV value (Fig. \ref{fig:MEV_cumprob_block}), and proposal frequency statistics (Fig. \ref{fig:prop_selection_cumprob_week}), allows us to quantify the expected weekly increase of validator-side latency optimization. As these variables are independent events, we can extract values from their respective distributions to calculate the anticipated weekly MEV increase — see Fig. \ref{fig:mev_inc_week_vp}.

As expected, the probability density function for the validator with lower voting power exhibits a wider spread, corresponding to comparatively higher variance in expected MEV increases. Conversely, the validator with a higher voting power exhibits a more concentrated distribution, corresponding to more frequent proposals, i.e. the total additional MEV payoff is comparatively more predictable.

\begin{figure}[h!]
  \centering
  \includegraphics[width=0.8\textwidth]{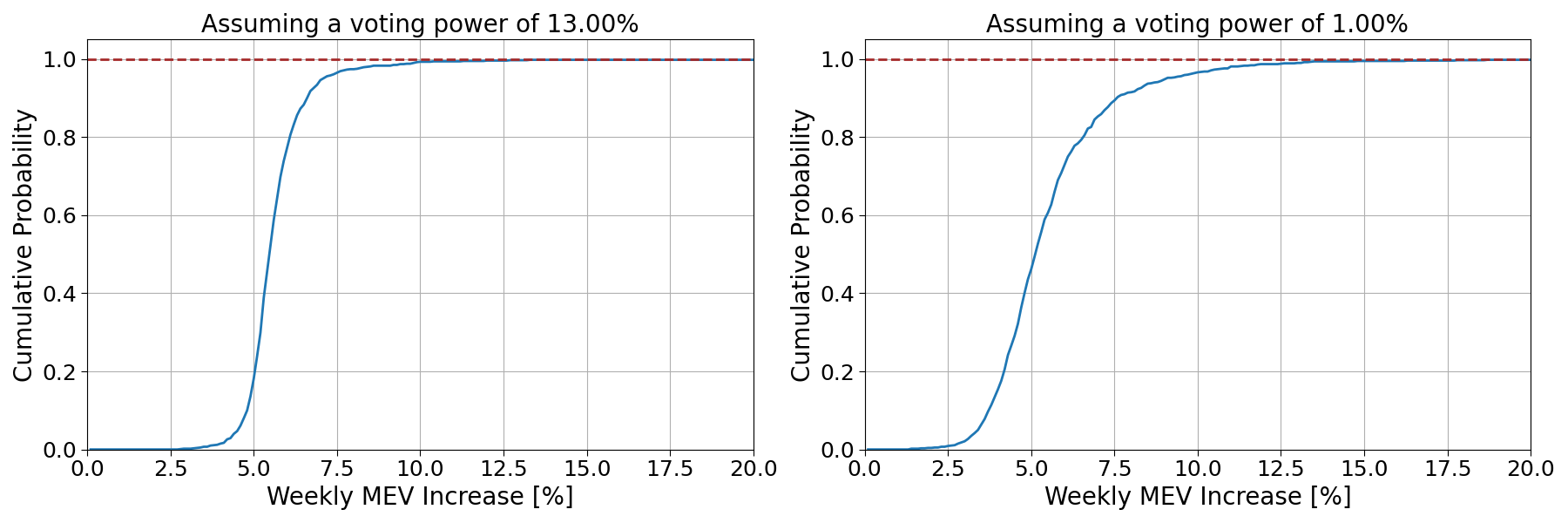}
  \caption{Cumulative probability of weekly MEV reward increases for a node operator with 13\% voting power (left panel) and 1\% voting power (right panel).}
  \label{fig:increase_cumprob_weekly}
\end{figure}

Figure \ref{fig:increase_cumprob_weekly} illustrates the cumulative probabilities of the percentage-wise MEV increase for each voting power scenario. Notably, the higher variance in the lower voting power scenario decreases the steepness of the cumulative probability curve. This is indicated by the broader spread in the corresponding probability density function. For a node operator with 13\% total voting power the median weekly MEV reward increase is at 5.47\%, with the lower quartile at 5.13\% and the upper quartile at 7.06\%. For a node operator with 1\% total voting power, the median increase is lower 5.11\%, but with a wider spread between the 25\%-quantile at 4.33\% and the 95\%-quantile at 9.03\%. The implication is that the return profiles for large- and small- node operators fit different client (delegator) utility functions as they are situated at different points of the risk / reward curve. Therefore, latency optimization is worthwhile for small- and large- node operators.

As the time horizon extends, variance for any voting power profile will naturally diminish due to the law of large numbers. This will likely concentrate rewards around the 5\% mark for any voting power profile. Overall, if execution layer rewards constitute roughly to 30\% of all rewards, a latency-aware MEV strategy can enhance APR from 4.2\% to 4.27\%. This corresponds to a significant 1.67\% uplift of overall APR. In conclusion, this is a significant margin, and node operators are incentivized to capture it.

\section{The cost of artificial latency}\label{sec:cost_of_latency}

In the previous section, we have seen how all institutional (i.e. client-facing) node operators are incentivized to compete for latency-optimized MEV capture, irrespective of their voting power. While the potential for equitable competition exists as different points on the risk / return curve, the introduction of an artificial delay into the block proposal process carries negative externalities which extend to subsequent proposers, and the network at large, rendering such a proposition naught. 

\subsection{Negative externalities due to an inflated ETH burn rate}

In the Sec. \ref{subsec:pbs_auct}, Fig. \ref{fig:bid_value_vs_gas_vs_eligibility} illustrates the upward trend in gas consumption as the auction progresses. This subsection will examine the direct correlation between the introduction of artificial latency to the auction and its impact on the Ethereum network’s gas dynamic; the discussion will focus on the fee burn mechanism introduced by EIP-1559 \cite{eip1559}.

\begin{figure}[h!]
  \centering
  \includegraphics[width=0.8\textwidth]{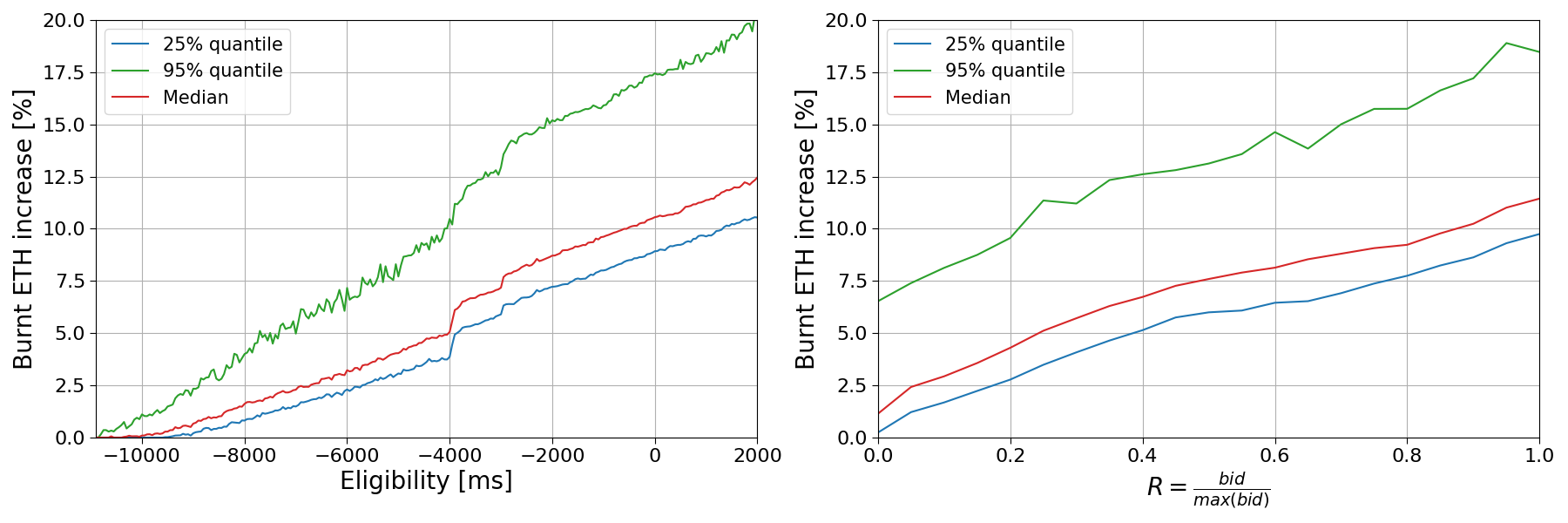}
  \caption{(Left panel) Burnt ETH increase for the subsequent block as function of the eligibility of the bid. (Right panel) Burnt ETH increases for the subsequent block as function of the bid value increase. For both panels, the red line represents the median of the distribution, the blue line represents the 25\%-quantile, and the green line represents the 95\%-quantile.}
  \label{fig:mev_network_cost}
\end{figure}

Figure \ref{fig:mev_network_cost} extends the previous visualization to map ETH burn against the bid eligibility time. The left panel illustrates how the percentage of burnt ETH for the next slot increases over time, and the right panel correlates the $R$ value with this percentage increase. The graph explicitly demonstrates that as bids rise during the auction, so does the gas price for the next slot, leading to a larger share of burned ETH in subsequent slots. The upshot is that artificial latency imposes a hidden cost on subsequent proposers, as a relatively larger share of their income is burned. While the base fee increases, if the opportunity for MEV extraction remains constant, builders are compelled to adjust the final portion of rewards they are willing to pay, effectively burning a part of what could have been the proposer's income. For normal transactions, the priority fee (PF) paid remains the same regardless of the base fee level, assuming the max fee isn't binding. But if the max fee is close to the base fee, an increase in base fee can lead to a reduction in the PF available for normal transactions, affecting again the income of the next proposer.

While sophisticated validators capture an upside from latency-optimized MEV capture, the systematic repercussions manifest as increased gas costs and an accelerated ETH burn rate for the subsequent proposers. The Ethereum network seeks to maximize decentralization by encouraging hobbyists to run validators. We demonstrate that these downside risks are significant in scale, and disproportionately impact solo validators.

\begin{figure}[h!]
  \centering
  \includegraphics[width=0.8\textwidth]{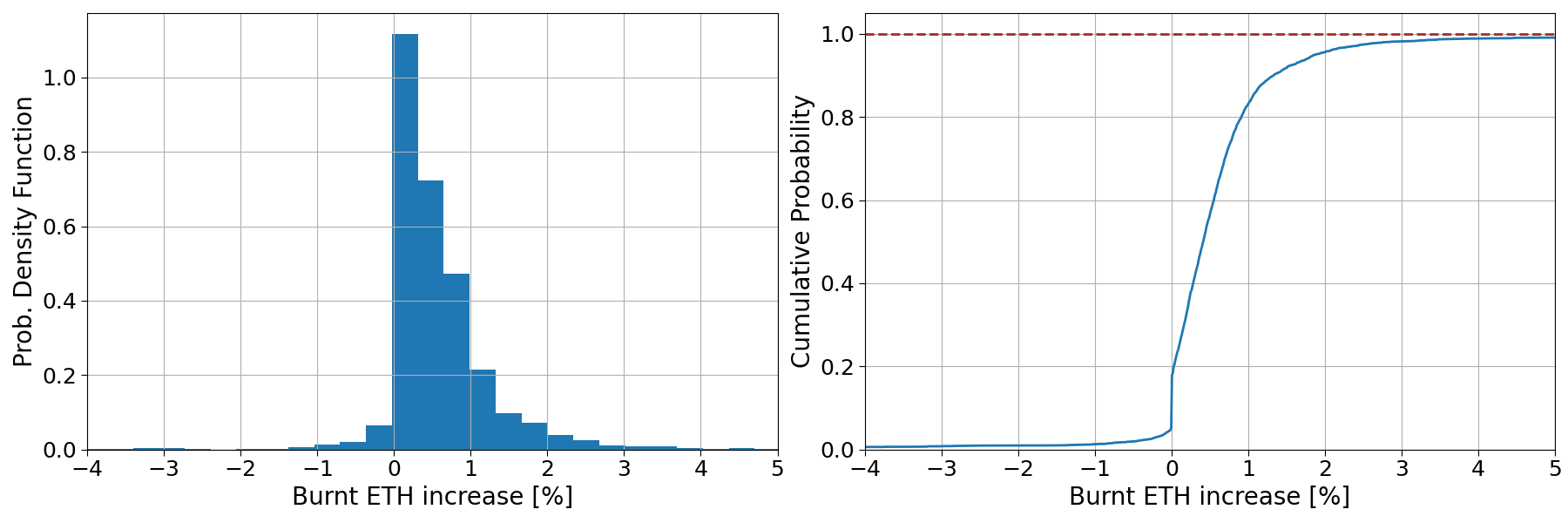}
  \caption{(Left panel) PDF of the burnt ETH increase obtained after applying the 950 ms standard delay. (Right panel) Cumulative probability of burnt ETH increase obtained after applying a delay.}
  \label{fig:increase_pdf_and_cumprob_burnt_eth_per_block}
\end{figure}

Figure \ref{fig:increase_pdf_and_cumprob_burnt_eth_per_block} demonstrates that the introduction of artificial latency into the auction increases the percentage of ETH burned meaningfully. The left panel displays the probability density function for the additional percentage of ETH burned. While the median increase of around 0.4\% appears modest, the breadth of the distribution indicates a wide spectrum of potential outcomes. Notably, even a small increase in burnt ETH can disproportionately reduce final rewards due to the typically larger amount of burnt ETH compared to MEV rewards.

A 0.5\% rise in burnt ETH translates into a tangible reduction in MEV rewards. For example, if an original MEV reward is 0.077 ETH with a burnt fee of 0.633 ETH, a 0.5\% increase in burnt ETH would lower the MEV reward to approximately 0.074 ETH. This reduction represents a subtle yet impactful 3.9\% decrease in the proposer's revenue. For node operators with relatively lower voting power, i.e. who are relatively less frequently chosen to propose blocks, the tail of the distribution poses a significant risk. Specifically, the 95\%-quantile settles around a 2\% increase in burnt fees, and higher values are possible. Consider that there is a 1\% probability that the increase exceeds 5\%. This can manifest a tangible effect on the overall APR of such a provider, rendering them non-competitive. 

This additional burn rate is consequently most impactful for solo validators, whose execution layer income would not only decrease, but be subject to greater variance. In short, the smaller a node operator, the more likely adverse impacts from latency games are to manifest, with hobbyist solo validators landing on the least desirable end of the risk / reward spectrum. The next subsection will extend the analysis to demonstrate that even sophisticated node operators compete in a zero sum game, for which the revenues captured from validators that do not optimize for latency serve as the seed capital. 

Before moving on, it is worth noting the presence of a negative tail in the probability density function shown in Fig. \ref{fig:increase_pdf_and_cumprob_burnt_eth_per_block}. This could be attributable to builders who place new bids not specifically to outcompete others, but to effectively supersede, i.e. cancel, a previously leading bid. This would be relevant when the opportunity the transaction in question seeks to exploit dissipates.

\subsection{A zero sum game for node operators}

\begin{figure}[h!]
  \centering
  \includegraphics[width=0.8\textwidth]{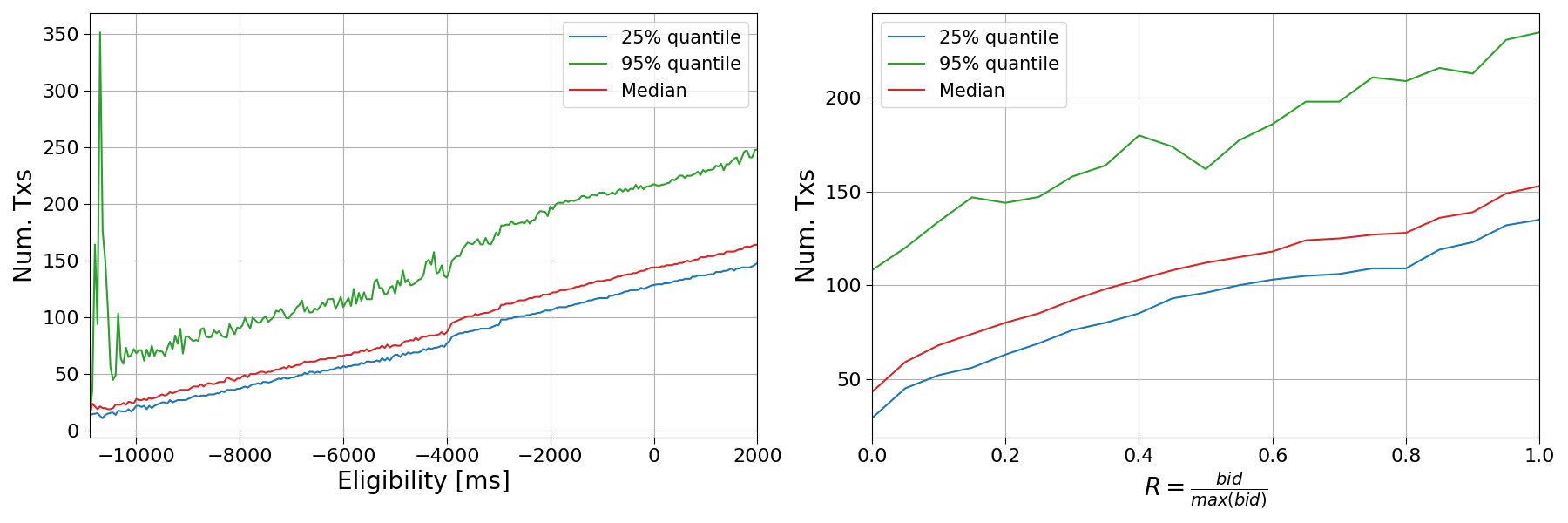}
  \caption{(Left panel) Dynamics of the number of transactions included in the block during the auction. (Right panel) Number of transactions included as a function of the increase in the bid value. For both panel, the red line represents the median of the distribution, the blue line represents the 25\%-quantile, and the green line represents the 95\%-quantile.}
  \label{fig:latency_vs_ntxs}
\end{figure}

\noindent
Figure \ref{fig:latency_vs_ntxs} illustrates how proposer behavior can alter the composition of a block, and consequently, the distribution of rewards. The left panel of the figure illustrates a clear trend: as proposers delay the \textbf{getHeader} request, there is a concomitant increase in the number of transactions included in a block. This increase is intuitive; more time allows for additional transactions to be pooled from the mempool into the block, potentially boosting the MEV available to the proposer. The right panel of the curve contextualizes this analysis further, demonstrating that as the reward value of a block nears its maximum ($R$ value approaching 1), the rate at which transactions are included rises (i.e. the slope of the curve). This suggests that in the latter stages of a slot, as new transactions continue to enter the mempool, builders have a larger opportunity space, increasing block value. 

This dynamic is twofold. First, vertically integrated builders (who are also searchers) can afford to place higher bids. As the time gap between centralized exchange (CEX) and decentralized exchange (DEX) settlement narrows, the price risk of inventory diminishes, allowing builders to bid more aggressively and widely. The next subsection will illustrate the downside impact of this in more detail. Secondly, this pattern implies that transactions that may have been included in the next slot land in the current one. This shifts potential MEV revenue from the future proposer to the incumbent, and gives rise to a zero sum game: the gain of one player is the loss of another.

\subsection{An increased LVR burden on LPs}

Liquidity providers (LPs) grapple with Loss-Versus-Rebalancing (LVR) \cite{LVR}, where arbitrageurs exploit stale prices to the detriment of LPs. The LVR metric captures the losses incurred by Automated Market Maker (AMM) LPs when their liquidity is traded against by arbitrageurs reacting to price movements between CEXs and DEXs.

LVR is sensitive to block times \cite{lvr_block_time}; higher durations exacerbate the information disconnect between venues and thus increase LP losses. The consequence is that delaying the getHeader request extend the opportunity window and decreases the risk for such arbitrage, imposing additional losses on LPs that would not be incurred under standard conditions. Research has consistently demonstrated a correlation between the first transactions in a block, and LP losses (see e.g. \cite{markoutsLVR, hedgLP}); these transactions generally involve cross-venue arbitrage. A successful arbitrageur has consistent access to the early slots in the block, and requires competitive execution on the CEX-side (i.e. fee tiers; infrastructure) to competitively bid.

In the first section of this study, we demonstrate that some builders profit from a consistent information advantage. This can materialize as the ability to execute CEX to DEX arbitrage competitively. The introduction of artificial latency by validators increases the range and profitability available to such a builder, at the expense of LPs, thus raising the aggregate cost of providing on-chain liquidity.

\subsection{An increase in centralization pressure}

The previous sections highlighted multiple ways dynamics intrinsic to PBS influence the Ethereum ecosystem. This section will synthesize the preceding discussion into specific ways these lead to centralization.

We examined how strategic delays by validators in submitting \textbf{getHeader} requests result in an increased ETH burn rate. This knock-on effect benefits node operators engaging in such timing games, to the detriment of others, that are net exposed to a higher base fee if proposing in subsequent slots. Additionally, node operators with a relatively lower voting power are exposed to disproportionately more variance from the long tail of percentage increases in ETH burned. In summary, large node operators playing timing games benefit from comparatively higher APR at lower variance to the detriment of other operators. We also examined how late block proposals require builders to include relatively more transactions to keep their bids competitive, thereby draining potential MEV profit from future blocks. This again manifests a disadvantage for smaller node operators, who propose blocks less frequently, and are therefore more exposed to individual block payoff variance. Finally, we highlighted how validator-side strategic timing games lead to higher LP losses from increased LVR, shifting profit to sophisticated CEX to DEX arbitrageurs, which can capitalize on more opportunities and bid more aggressively due to decreased inventory risk. A share of the direct upside again accrues to latency optimized node operators, reinforcing centralization pressure. Additionally, LPs may diversify their capital deployment to include a mix of liquidity provisioning and hedging through MEV-optimized staking \cite{hedgLP}, manifesting further centralization pressure.

Across the board, the nature of MEV favors node operators with relatively higher voting power, who naturally capture returns at lower variance. Strategic latency games compound this effect in the ways highlighted above, potentially manifesting risks for the Ethereum network at large due to increased centralization, and potentially higher gas fees. Additionally, large node operators enjoy access to a larger pool of in-client data (e.g. bid timings from MEV-Boost), which allows more efficient hypothesis testing, and reflects as more effective latency parameters. This edge scales with voting power.

Within this context, node operators are compelled to employ latency optimization as a matter of strategic necessity. As more node operators exploit these inefficiencies, they progressively increase the benchmark rate for returns, giving capital providers a simple heuristic via which to select for latency-optimized setups. This further perpetuates the latency game, ossifying it into standard practice, upping the pressure on node operators reluctant to participate, in a self-reinforcing cycle. Ultimately, this manifests as an environment where a node operator’s competitive edge is defined by its willingness to exploit this systematic inefficiency.

\section{Empirical results from the Adagio pilot}\label{sec:adagio_results}

In late August 2023, Chorus One \cite{c1web} launched a latency-optimized setup — internally dubbed \textit{Adagio} on Ethereum mainnet. It’s goal was to gather actionable data in a sane manner, minimizing any potential disruptions to the network. Until this point, \textit{Adagio} has not been a client-facing product, but an internal research initiative running on approximately 100 self-funded validators. We are committed to both operational honesty and rational competition, and are therefore disclosing our findings via this study.

Our pilot comprises four distinct setups, each representing a variable (i.e. a relay) in our experiment:
\begin{itemize}
    \item \textbf{The Benchmark Setup:} Two relays operate without latency modifications, serving as a control group. These allow us to measure the impact of our experimental variables against a vanilla setup, ensuring a baseline for comparison. Both relays are non-optimistic. Further, these function as a safety net in case the artificially delayed relays fail to deliver blocks on time. 
    \item \textbf{The Aggressive Setup:} This approach pushes the boundaries of the auction's timing, delaying the \textbf{getHeader} request as much as reasonably possible on a risk-adjusted basis (i.e. to the brink of the auction's temporal limit). It is designed to capture the maximum possible MEV. This setup is features a non-optimistic relay.
    \item \textbf{The Normal Setup:} This setup rationally balances MEV capture against potential adverse impacts, and only delays the auction within a ``safe'' parameter space. It carries a favorable risk-reward profile. This setup features a optimistic relay.
    \item \textbf{The Moderate Setup:} This is our most conservative setup; it terminates the auction slightly ahead of our estimated safety threshold (100 ms prior), minimizing risk to the network while still engaging in competitive optimization. This setup features a optimistic relay.
\end{itemize}

The Adagio pilot is an exploration -- its purpose extends beyond understanding how varying degrees of latency optimization affect both our aggregate profitability and the network at large. It also aims to mitigate the risk of bid cancellation \cite{bidCanc} when the same block is sent to multiple relays. By employing different latency setups, we can sample in distinct time regions, effectively preventing overlap and ensuring a more robust and efficient bidding process. If successful, it balances between rational self-interest as a node operators with our responsibility to the network's integrity and performance, i.e. exists within the context of decentralized incentive alignment.

\subsection{Results}

In this section, we present a comprehensive outcome analysis for our \textit{Adagio} pilot. Our primary objective is to examine the influence of different relay configurations on the timing of bid selection and eligibility; this is central to the dynamics of the MEV-Boost auction.

\begin{figure}[h!]
  \centering
  \includegraphics[width=0.8\textwidth]{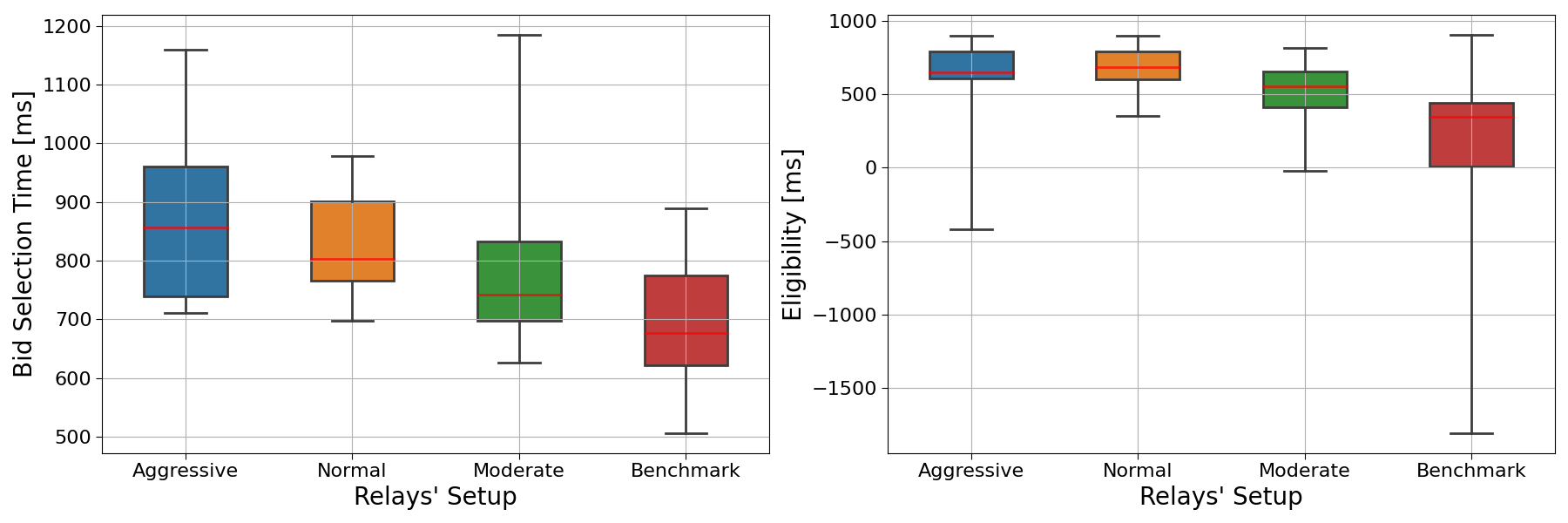}
  \caption{(Left panel) Box plot of the bid selection time from the relays with different setup. The time is with respect the  slot time. (Right panel) Box plot of the eligibility of received bid from relays with different setup. For both panels, the red lines represent the medians of the distributions, meanwhile the boxes represent the distributions between the 25\% and 75\% quantiles.}
  \label{fig:relays_setup_vs_elig}
\end{figure}

First, we examine how the intrinsic latency of each relay shapes the overall auction dynamic. Figure \ref{fig:relays_setup_vs_elig} shows the timestamps of bid selection, and the corresponding bid eligibility times. Notably, the graphs highlight instances where the relay response times do not correspond to the expected latency, e.g. in the case of the normal setup's unexpectedly quicker responses versus the aggressive setup. This can be indicative of performance differences between relays; a future study may analyze this further in the context of timely payload delivery. 

Irrespectively, the data confirms that our parameters generally and safely insure bid selection within the 1-second mark of the slot. This operational threshold is central, as it minimizes the risk of network congestion, and the risk of forks caused by late block propagation. On the operator side, it also minimizes the risk of a 0 payoff via a missed slot.

We proceed with an analysis of each setup's behavior:
\begin{itemize}
    \item The \textbf{Benchmark Setup} adheres to expected performance standards, with median bid eligibility occurring at 344.5 milliseconds into the slot, and the 95\% quantile at 575.75 milliseconds. This seems to indicate that the intrinsic latency (i.e. network topology or relay idiosyncrasies) can organically mirror artificial latency. Our benchmark setup seems to select bids in the right tail of the network distribution (see Fig. \ref{fig:eligibility_pdf_and_cumprob}).
    \item The \textbf{Aggressive and Normal Setups} exhibit competitive bid eligibility timings, with the aggressive setup surprisingly under-performing the normal setup despite its higher latency parameter. This is indicative of inherent differences between relays, such as the time to process bids, and their geographical placement.
    \item The \textbf{Moderate Setup} outperforms the benchmark in terms of delay but lags behind the aggressive and normal setups. This result is particularly interesting as it indicates that non-optimistic relays could be introducing an artificial latency parameter to remain competitive with the optimistic relays.
\end{itemize}

In summary, the data suggests that latency strategies within relay operations carry significant implications for relay competitiveness. The aggressive setup in particular appears to enable non-optimistic relays to operate on par with optimistic peers. The practical upshot is that some relays can only consistently compete through an artificial delay. An extreme case of this would be a relay which is technically consistently non-competitive, but captures exclusive order flow -- in this case, a rational node operator will always query it with an artificial latency parameter. This competition among relays hints at a potential evolutionary shift in the game, where relays might strategically delay responses to enhance their chances of providing the best bid. This adaptation could be likely incentivized by the benefits of optimized delay, particularly for vertically-integrated builders who may be inclined to pay a premium for more reliable bid inclusion.

These results provide valuable insights into how strategic latency implementation within the relay infrastructure could influence the aggregate efficacy and competitive landscape of the MEV-Boost auction, by leveling the playing field between different relays via custom latency parameter.

\begin{figure}[h!]
  \centering
  \includegraphics[width=0.8\textwidth]{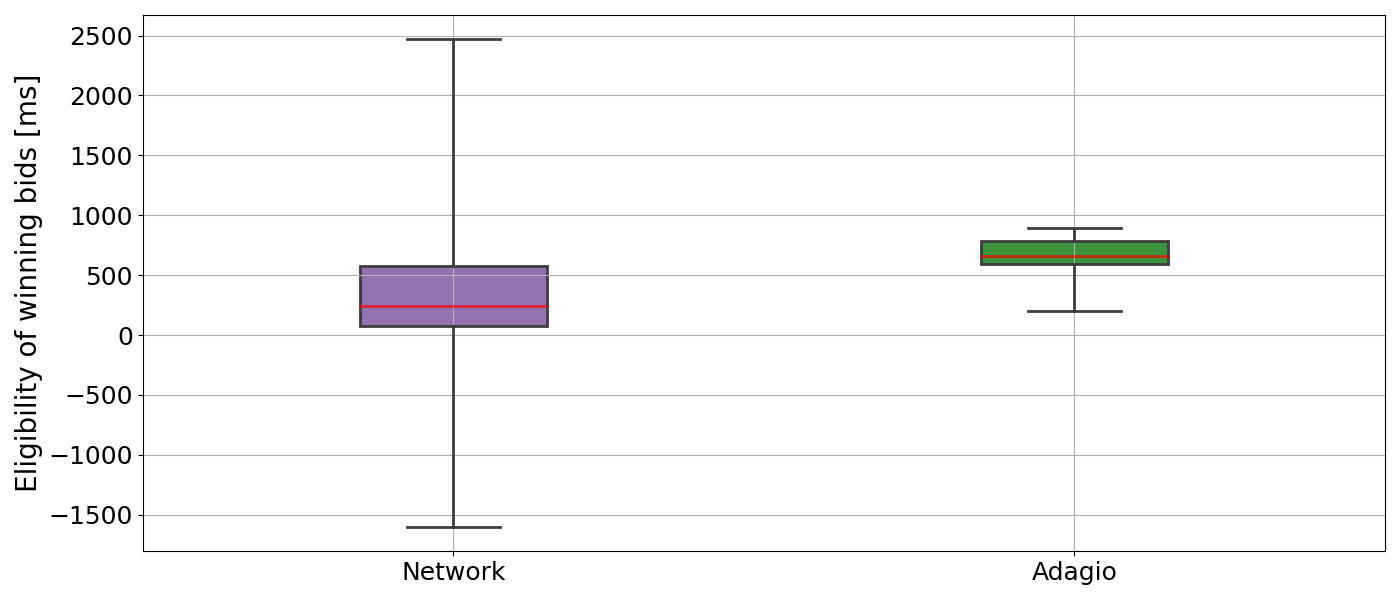}
  \caption{Box plot of the eligibility time of winning bids. The red lines represent the medians of the distributions, meanwhile the boxes represent the distributions between the 25\% and 75\% quantiles.}
  \label{fig:eligibility_c1_vs_network}
\end{figure}

Figure \ref{fig:eligibility_c1_vs_network} shows the eligibility of winning bids for Adagio, compared with the network distribution. Despite the heterogeneous latency settings for relay setups, the data reveals a consistent pattern: our strategy predominantly selects bids that become eligible after 500 milliseconds, with the 25\%-quantile marked at 589.5 ms. This consistency highlights the Normal and Aggressive setups as the primary contributors to winning bids, with a median selection time of 656.0 ms and a 95\%-quantile at 886.35 ms.

Critically, these findings underscore our decision to sample below the 950 ms threshold posited in our theoretical framework; instead, we aimed for an average delay of 700 ms. There are two reasons for this. First, in a realistic setting, sampling around the 950ms mark can cause selections beyond the 1s mark due a given relay's inherent latency, or network topology. Operating without a safety margin risks congestion and forks via late blocks, and increases the risk of a missed slot. Second, there is insufficient statistical evidence to justify pushing beyond this threshold. As the bid increase flattens out after 950 ms (see Fig. \ref{fig:bid_value_vs_gas_vs_eligibility}), i.e. warps the risk / reward ratio unfavorably. To be exact, while the median value for the ratio defined in Eq. \ref{eq:Restimator} increased by 3.39\% from 250 ms to 950 ms, it increased by only 0.18\% between 950 ms and 1 s.

We will conclude by estimating the profit uplift realized by Adagio. In our theoretical framework, we assumed a static delay (see. Fig. \ref{fig:increase_pdf_and_cumprob_real}); in practice, we observe a fluctuation in the eligibility time distribution. To offset this, we will sample eligibility times from the cumulative distribution in Fig.3, and compute how much additional MEV revenue has likely been realized for an eligibility delay corresponding to our actual observations (i.e. see Fig. \ref{fig:eligibility_c1_vs_network}).

\begin{figure}[h!]
  \centering
  \includegraphics[width=0.8\textwidth]{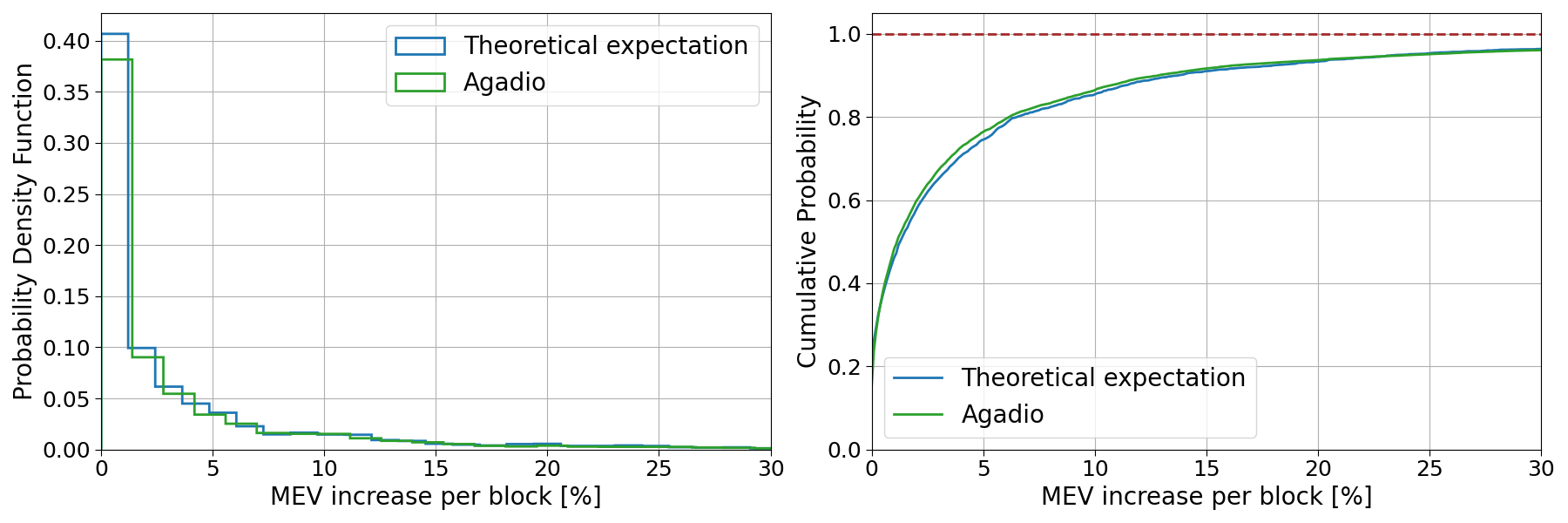}
  \caption{(Left panel) Probability density function of MEV increases per block. (Right panel) Cumulative probability of MEV increases per block. In both panels, the blue line represents the theoretical model obtained assuming we always hit the 950 ms value for the eligibility time of bids. The green line represents the expectation obtained using the Adagio data.}
  \label{fig:increase_pdf_and_cumprob_adagio_vs_optimal}
\end{figure}

Figure \ref{fig:increase_pdf_and_cumprob_adagio_vs_optimal} plots the result of these simulations, comparing the theoretically expected MEV revenue against the empirical data from the Adagio pilot. The probability density function shows a wider variance with a higher cumulative probability. This indicates that while the experiment leads to a broader range of outcomes, it skews towards relatively lower MEV increases in the practical setting. Combining the latency optimization payoff (Fig. \ref{fig:increase_pdf_and_cumprob_adagio_vs_optimal}), per-block MEV value (Fig. \ref{fig:MEV_cumprob_block}), and proposal frequency statistics (i.e. using the Adagio VP), allows us to quantify the expected annual increase of validator-side latency optimization (i.e. empirically mirroring Fig. \ref{fig:mev_inc_week_vp}). 

\begin{figure}[h!]
  \centering
  \includegraphics[width=0.8\textwidth]{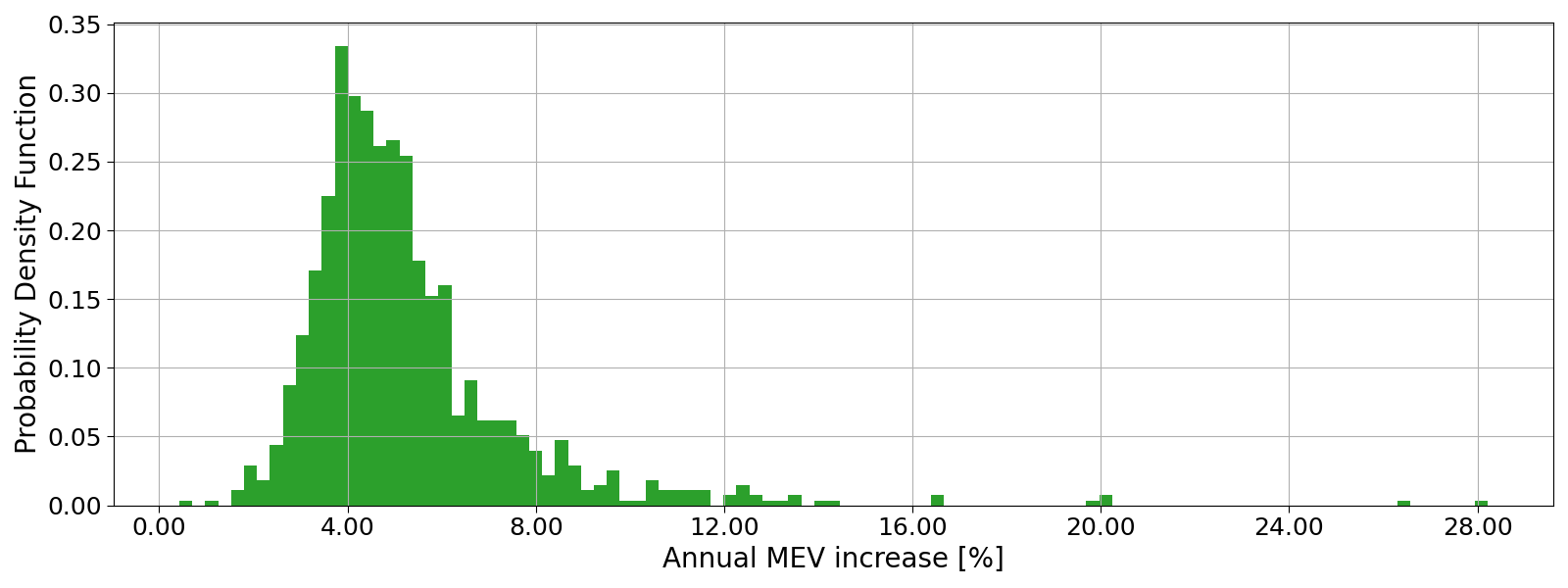}
  \caption{PDF of annual MEV increase expected by adopting the Adagio setup. The high spread is due to the low voting power we have with the current pilot.}
  \label{fig:annual_mev_inc_adagio}
\end{figure}

The simulation results plotted in Fig. \ref{fig:annual_mev_inc_adagio} indicate a median MEV increase per year at 4.75\%, with the interquartile range extending from 3.92\% to 9.27\%. This correspond to an APR 1.58\% higher than the vanilla case, with interquartile range from 1.30\% to 3.09\%. The increased spread primarily arises from the pilot's constrained voting power. However, a portion of it is due to a fluctuation in bids eligibility. Further, the observed median comes in 5\% than the theoretical projection. To bridge this gap, we will update our approach so as to minimize variance in bid selections, and maintain eligibility times below the 950 ms threshold.

\section{Conclusions}

This study has examined how strategic timing games in the PBS context take advantage of the dynamics of the MEV-Boost auction to yield additional MEV revenue for node operators. In this context, it discussed node operator incentives, and highlighted how the externalities of such latency optimization create systemic challenges for the Ethereum network. Specifically, these include node operator centralization and associated risks, network inefficiencies including potentially higher gas prices, and an increased LVR burden on LPs. 

We illustrated that all node operators are incentivized to compete for latency-optimized MEV capture, irrespective of their voting power. While in principle, the potential for equitable competition between node operators of different size exists as different points on the risk / return curve (i.e. variance), the introduction of artificial latency carries negative externalities which affect subsequent proposers, rendering such a proposition naught.

Further, we argued that, as an increasing share of node operators employs latency optimization, these progressively increase the benchmark rate for returns heuristic via latency-optimized setups. In this way, the opportunity cost of not engaging in timing games increases in a self-reinforcing cycle. Ultimately, this normalizes such optimization as standard operating procedure, and manifests an environment in which any node operator’s competitive edge is defined by its willingness and ability to exploit systematic inefficiencies.

We highlight how strategic timing games lead to higher LP losses from increased LVR, benefiting statistical arbitrageurs. As the auction duration extends, these can capitalize on more opportunities due to decreased inventory risk and a larger disconnect between CEX and DEX pricing. A profit share directly accrues to latency optimized node operators; additionally, LPs may stake with such node operators to offset some of their LVR risk, manifesting further centralization pressure. 

We argued that, large node operators enjoy a systematic edge through reduced payoff variance via a higher block proposal frequency, and access to a larger pool of in-client data that reflects as more effective latency parameters. We demonstrated that artificial latency can result in a higher base fee, and that the long tail of the percentage increase in ETH burned poses a significant risk to small node operators and hobbyist validators. 

Lastly, we presented our \textit{Adagio} pilot and demonstrated that latency optimization significantly increases node operator revenue, at an estimated 1.58\% boost to APR versus a non-optimized setup. We illustrated how practical latency parameters can balance competitiveness with network health, and thus provide insights to smaller node operators which may struggle to find sufficient data to stay competitive. Our research emphasizes the need to take a cautious and informed approach to latency optimization, rationally weighing competitive need versus potential drawbacks. 

Overall, this study provides insight into node operator incentives, the dynamics of the MEV-Boost auction, and the cost of artificial latency in a PBS context. Future work may examine specific centralization risks in depth, such as cross-block MEV, and contribute to specific mitigation strategies, such as MEV-burn.

\section*{Acknowledgments}

\noindent
This research has been granted by Chorus One. We are grateful to G. Sofia, B. Crain, and F. Lutsch for useful discussions and comments. We acknowledge support from the Chorus One enginering team for the implementation of the MEV-Boost changes. We also thank B. Monnot and C. Schwarz-Shilling for the rewiev of the entire document.

\bibliographystyle{plain}
\bibliography{sample}

\begin{thebibliography}{10}

\bibitem{lvr_block_time}
@0x94305.
\newblock Lvr dependencies on block time.
\newblock \url{https://x.com/0x94305/status/1577683141346459648?s=20}.

\bibitem{bidSubRelay}
Relays’ API.
\newblock Builder bids submission.
\newblock \url{https://flashbots.github.io/relay-specs/#/Data/getReceivedBids}.

\bibitem{optRel}
A.~Chiplunkar and M.~Neuder.
\newblock Optimistic relays and where to find them.
\newblock \url{https://frontier.tech/optimistic-relays-and-where-to-find-them}.

\bibitem{markoutsLVR}
CrocSwap.
\newblock Usage of markout to calculate lp profitability in uniswap v3.
\newblock \url{https://crocswap.medium.com/usage-of-markout-to-calculate-lp-profitability-in-uniswap-v3-e32773b1a88e}.

\bibitem{MEV}
P.~Daian, S.~Goldfeder, T.~Kell, Y.~Li, X.~Zhao, I.~Bentov, L.~Breidenbach, and A.~Juels.
\newblock Flash boys 2.0: Frontrunning, transaction reordering, and consensus instability in decentralized exchanges.
\newblock \url{https://arxiv.org/pdf/1904.05234v1.pdf}.

\bibitem{data_always_bid_time}
@Data\_Always.
\newblock Distribution of winning bid arrival time.
\newblock \url{https://twitter.com/Data_Always/status/1734686026579493345}.

\bibitem{pbs_specs}
Ethereum.
\newblock Builder specs (validator).
\newblock \url{https://github.com/ethereum/builder-specs/blob/main/specs/bellatrix/validator.md}.

\bibitem{eip1559}
Ethereum.
\newblock eip-1559.md.
\newblock \url{https://github.com/ethereum/EIPs/blob/master/EIPS/eip-1559.md}.

\bibitem{mevboost_specs}
Flashbots.
\newblock Mev-boost specs.
\newblock \url{https://github.com/flashbots/mev-boost/blob/6356e799bdfb5b11a35d6ec3ec70712c79c02e20/cli/main.go}.

\bibitem{lowcarbcruc}
Flashbots.
\newblock Post mortem: April 3rd, 2023 mev-boost relay incident and related timing issue.
\newblock \url{https://collective.flashbots.net/t/post-mortem-april-3rd-2023-mev-boost-relay-incident-and-related-timing-issue/1540}.

\bibitem{optFeeMarket}
S.~Leonardos, D.~Reijsbergen, B.~Monnot, and G.~Piliouras.
\newblock Optimality despite chaos in fee markets.
\newblock \url{https://arxiv.org/pdf/2212.07175.pdf}.

\bibitem{LVR}
J.~Milionis, C.~Moallemi, T.~Roughgarden, and A.~Lee Zhang.
\newblock Automated market making and loss-versus-rebalancing.
\newblock \url{https://arxiv.org/pdf/2208.06046.pdf}.

\bibitem{MEVhealtDash}
U.~Natale.
\newblock Ethereum mev healthiness dashboard.
\newblock \url{https://flipsidecrypto.xyz/edit/queries/951251b7-757b-46b1-aee5-50b80f1f96cf}.

\bibitem{hedgLP}
U.~Natale.
\newblock Hedging lp position by staking.
\newblock \url{https://chorusone.notion.site/Hedging-LP-position-by-staking-651f4e543f2448a58d99b788d35d941b?pvs=4}.

\bibitem{bidCanc}
M.~Neuder.
\newblock Bid cancellations considered harmful.
\newblock \url{https://ethresear.ch/t/bid-cancellations-considered-harmful/15500}.

\bibitem{c1web}
Chorus One.
\newblock Chorus one webpage.
\newblock \url{https://chorus.one/}.

\bibitem{time_is_money}
C.~Schwarz-Schilling, F.~Saleh, T.~Thiery, J.~Pan, N.~Shah, and B.~Monnot.
\newblock Time is money: Strategic timing games in proof-of-stake protocols.
\newblock \url{https://arxiv.org/pdf/2305.09032.pdf}.

\bibitem{MEVBoostDash}
T.~Wahrst\:atter.
\newblock Mev-boost dashboard.
\newblock \url{https://mevboost.pics/}.

\end{thebibliography}

\end{document}